\documentclass[reprint,twocolumn,amsmath,amssymb,aps,pra,longbibliography,floatfix]{revtex4-1}
\usepackage{physics}
\usepackage{siunitx}
\usepackage{amsmath}
\usepackage{cases}
\usepackage{url}
\usepackage{natbib}
\usepackage{textcase}
\usepackage{amssymb}
\usepackage{graphicx}
\usepackage{bm} 

\usepackage{times}
\usepackage{float}
\usepackage{multirow,microtype,color,relsize}
\usepackage{subfigure}
\usepackage[breaklinks=true]{hyperref}
\usepackage[utf8]{inputenc}
\usepackage[english]{babel}
\hypersetup{colorlinks=true,linkcolor=blue,citecolor=blue}
\hypersetup{linktocpage}
\usepackage{CJKutf8}
\usepackage{breakcites}
\usepackage{float}
\usepackage[dvipsnames]{xcolor}
\definecolor{mypine}{RGB}{1, 121, 111}

\newcommand{\rref}[1]{Ref.\,\onlinecite{#1}}

\def \be {\begin{equation}}
\def \ee {\end{equation}}
\begin{document}
\begin{CJK*}{UTF8}{gbsn}
\title{Scattering mechanisms in state-of-the-art GaAs/AlGaAs quantum wells}

\author{Yi Huang~(黄奕)}
\email[Corresponding author: ]{huan1756@umn.edu}

\author{B.\,I. Shklovskii}

\author{M.\,A. Zudov}
 
\affiliation{School of Physics and Astronomy, University of Minnesota, Minneapolis, Minnesota 55455, USA}
\date{\today}

\begin{abstract}
Motivated by recent breakthrough in molecular beam epitaxy of GaAs/AlGaAs quantum wells [Y. J. Chung \textit{et al.}, Nature
Materials \textbf{20}, 632 (2021)], we examine contributions to mobility and quantum mobility from various scattering mechanisms and their dependencies on the electron density.
We find that at lower electron densities, $n_e \lesssim 1 \times 10^{11}$ cm$^{-2}$, both transport and quantum mobility are limited by unintentional background impurities and follow a power law dependence, $\propto n_e^{\alpha}$, with $\alpha \approx 0.85$. 
Our predictions for quantum mobility are in reasonable agreement with an estimate obtained from the resistivity at filling factor $\nu= 1/2$ in a sample of Y. J. Chung \textit{et al.} with $n_e = 1 \times 10^{11}$ cm$^{-2}$. 
Consideration of other scattering mechanisms indicates that interface roughness (remote donors) is a likely limiting factor of transport (quantum) mobility at higher electron densities. 
Future measurements of quantum mobility should yield information on the distribution of background impurities in GaAs and AlGaAs.
\end{abstract}

\maketitle
\end{CJK*}

Advances in molecular beam epitaxy, particularly the purification of source materials and improved vacuum conditions, have recently lead to a new generation of GaAs/AlGaAs quantum wells \cite{chung:2021} in which the mobility reached a record value of $\mu  = 4.4 \times 10^7$ cm$^2$V$^{-1}$s$^{-1}$ at electron density $n_e = 2.0 \times 10^{11}$ cm$^{-2}$.
The increase in mobility was especially pronounced at lower densities ($n_e \lesssim 1 \times 10^{11}$ cm$^{-2}$), where $\mu$ was twice the previously reported record values.
At higher densities ($n_e \gtrsim 2 \times 10^{11}$ cm$^{-2}$), however, the mobility has decreased to $\mu  \approx 3.5 \times 10^7$ cm$^2$V$^{-1}$s$^{-1}$ at $n_e \approx 2.7 \times 10^{11}$ cm$^{-2}$, approaching previously reported values.
While the increase of $\mu$ at low $n_e$ could be readily attributed to a reduced concentration of unintentional background impurities, subsequent reduction of $\mu$ at higher $n_e$ calls for revisiting other scattering sources.

In this Letter, we theoretically examine both transport and quantum mobility ($\mu_{\rm q}$) considering scattering by background impurities (BI), remote ionized donors in the doping layers (RI), interface roughness (IR), and alloy disorder (AD).
We find that both $\mu$ and $\mu_{\rm q}$ are limited by BI scattering at low $n_e$, as expected. 
With increasing $n_e$, however, scattering on IR (and eventually on AD) becomes important, as far as $\mu$ is concerned, whereas $\mu_{\rm q}$ becomes limited by RI scattering \citep{hwang}.

Modern GaAs-based heterostructures, such as those reported in \rref{chung:2021}, consist of a GaAs quantum well of width $w$ surrounded by Al$_x$Ga$_{1-x}$As barriers of thickness $d$. 
A two-dimensional electron gas (2DEG) with a concentration $n_e$ is supplied by two remote doping layers, each located at a distance $d_w = d + w/2$ from the center of the GaAs quantum well. 
These layers have a sophisticated doping well design, which helps to substantially reduce the scattering by ionized donors owing to excess electron screening \cite{sammon:2018,chung:2020a,akiho:2021,akiho:2022}. 
In this design, a $\delta-$layer of silicon atoms with concentration $n \gg n_e$ is implanted into a narrow ($\approx 3$ nm) GaAs quantum well, which is sandwiched between thin ($\approx 2$ nm) AlAs layers \cite{dw2}.
In our calculations we use $n \approx 1.5 \times 10^{12}$ cm$^{-2}$, Al mole fraction $x = 0.12$, and take into account the electron density dependencies of the effective spacer thickness $d_w$ and of the quantum well width $w$.
More specifically, we use $k_F w = 3.9$, where  $k_F = \sqrt{2\pi n_e}$ is the Fermi wave number, and $d_w^{-1} =  a n_e$, where $a = 3.55$ nm \citep{wdw}.
These constraints were obtained by fitting samples parameters of \rref{chung:2021}, as detailed in the Supplemental Material \citep{note:sm}.

We start with transport mobility, $\mu = e \tau / m^{\star}$, where $m^\star = 0.067\, m_0$ is the effective mass of an electron in GaAs~\citep{mass}, $m_0$ is the free electron mass, 
\begin{align}\label{eq:1_tau}
	\frac{1}{\tau} =\frac{4m^{\star}}{\pi \hbar^3} \int \limits_0^{2k_F}\frac{dq}{\sqrt{4k_F^2 - q^2}} \qty(\frac{q}{2k_F})^2 \ev{\abs{U(q)}^2}\,,
\end{align}
is the transport scattering rate, and $U(q)$ is the screened potential of a given scattering source.

\emph{Coulomb background impurities.}
\label{sec:background}
The screened potential squared averaged over impurity positions is given by 
\begin{align}\label{eq:uq2}
    \ev{\abs{U_{\rm BI}(q)}^2} = \int \limits_{-\infty}^{+\infty} dz \; N(z) U_1^2(q,z)\,,
\end{align}
where $U_1^2(q,z)$ is the screened potential squared from a single Coulomb impurity, and $N(z)$ is the 3D concentration of impurities at a distance $z$ from the center of the 2DEG.
Since the main contribution to momentum relaxation comes from impurities located close to the quantum well, for which excess electron screening \citep{sammon:2018} is not important and $U_1(q,z)$ can be obtained taking into account screening by electrons in the quantum well only.
Using Thomas-Fermi (TF) approximation~\citep{sammon:2018}, and following Refs.~\onlinecite{jonson:1976,gold:1986,gold:1987,gold:1989b,gold:2013} , we can write (more detailed discussion can be found in the Supplemental Material \citep{note:sm})
\begin{align}\label{eq:u_i1_mu}
    U_1(q,z) = \frac{2\pi e^2}{\kappa q \epsilon(q) }
    \int \limits_{-\infty}^{+\infty} dz' \;\abs{\psi(z')}^2 e^{-q\abs{z- z'}}\,,
\end{align}
where $\psi(z)$ is the wave function along $z$ direction and the dielectric function is given by
\begin{align}\label{eq:df}
    \epsilon(q) = 1 + (q_{\rm TF}/q) F_c (qw) [1- G(q)]\,.
\end{align}
Here, $q_{\rm TF} = 2/a_B$, $a_B = \kappa \hbar^2 / m^{\star} e^2$ is the effective Bohr radius, $\kappa = 12.9$ is the dielectric constant of GaAs, $G(q) = q/(2\sqrt{q^2 + k_F^2})$ is the local field correction using Hubbard approximation~\cite{jonson:1976}, and the form factor $F_c(qw)$ is given by
\begin{equation}\label{eq:fcq}
    F_c(qw) = \iint \limits_{-\infty}^{+\infty} dz dz' \abs{\psi(z)}^2  \abs{\psi(z')}^2 \exp(-q\abs{z-z'})\,.
\end{equation}
For small concentration $n_e < 1 \times 10^{11}$ cm$^{-2}$, it suffices to use an infinite-potential-well approximation and $\psi(z) = \sqrt{2/w} \cos{(z \pi/w)} \Theta(w/2 - \abs{z})$ in both Eq.~\eqref{eq:u_i1_mu} and Eq.~\eqref{eq:fcq}.

\begin{figure}
    \centering
    \includegraphics[width = \linewidth]{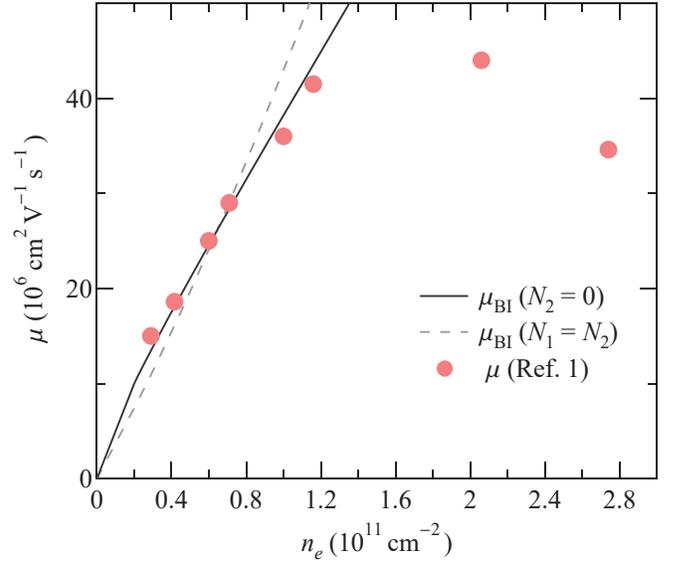}
    \caption{Mobility $\mu$ versus electron density $n_e$. 
   Circles are experimental data from Ref.~\onlinecite{chung:2021}. 
   Solid (dashed) line represents $\mu_{\rm BI}$ calculated for $N_1 = 1.4\times 10^{14}$ cm$^{-3}$, $N_2 = 0$ ($N_1 = N_2 = 5 \times 10^{12}$ cm$^{-3}$). 
   }
    \label{fig:mobility_low_ne}
\end{figure}

Following Ref.~\onlinecite{sammon:2018}, we define BI density as
\begin{align}
    N(z) = 
    \begin{cases}
        N_1, \; & w/2 < \abs{z} <d_w\, ,\\
        N_2, \; & \abs{z} < w/2\,,
    \end{cases}
\end{align}
where $N_1$ ($N_2$) represents the 3D concentration of impurities in AlGaAs (GaAs).
The results of our calculations for uniform impurity distribution ($N_1=N_2$) and for no impurities in GaAs ($N_2 = 0$) are shown in Fig.~\ref{fig:mobility_low_ne} along with the experimental data of \rref{chung:2021} (circles). 
We find that the data at $n_e \lesssim 1 \times 10^{11}$ cm$^{-2}$ can be better described by $N_1 = 1.4 \times 10^{14}$ cm$^{-3}$ and $N_2=0$ (solid line). 
Assuming uniform distribution of impurities (dashed curve) yields $N_1 = N_2 = 5 \times 10^{12}$ cm$^{-3}$, which is close to an estimate $1 \times 10^{13}$ cm$^{-3}$ of Ref.~\onlinecite{chung:2021}
~\footnote{This uniform impurities concentration is also close to a recent estimate found in Ref.~\onlinecite{ahn:2022}. However, they used fixed quantum well width $w = 30$ nm in the calculations, while we take into account the density dependence such that $k_F w = 3.9$.}.
We also note that $\mu_{\rm BI}(N_2=0)$ is well described by $\mu_{\rm BI} = 38.3\,n_e^\alpha$, where $\alpha \approx 0.85$, whereas $\mu_{\rm BI}(N_1=N_2)$  follows $\mu_{\rm BI} = 43.1\,n_e^\alpha$, where $\alpha \approx 1.12$.
Here and in what follows we assume that the electron density is in units of $10^{11}$ cm$^{-2}$ and  the mobility is in units of $10^6$ cm$^{2}$V$^{-1}$s$^{-1}$.

While BI scattering can describe the experimental data reasonably well at $n_e \le 1$, it is clear that BI scattering alone cannot explain experimental $\mu$ at higher $n_e$.
Indeed, obtained power laws represent crossovers from $\mu_{\rm BI}  \propto n_e^{1/2}$ at low $n_e$~\cite{sammon:2018} to $\mu_{\rm BI}  \propto n_e^{3/2}$ at higher $n_e$~\cite{dmitriev:2012}.
To explain the experimental $\mu$ at higher $n_e$, one needs to examine other scattering sources.

\emph{Remote impurities.}
One obvious candidate for reduced $\mu$ at higher $n_e$ is remote ionized impurity scattering.
For electron densities  $n_e < 10$ and for a given doping concentration $n = 1.5 \times 10^{12}$ cm$^{-2}$, the fraction of ionized donors in each doping layer is small, $1 - f = n_e/2n < 0.5$, and we can use Eq.~(22) from \rref{sammon:2018}~\cite{remote},
\begin{align}\label{eq:mur}
    \mu_{\rm RI} &= 7.7\,\frac{e}{\hbar} \frac{n k_F^3 d_w^5}{n_e} = 1.6 \times 10^6\, n_e^{-9/2}\,,
\end{align}
where in the final expression of Eq.~\eqref{eq:mur} we used $n=15$, $k_F = \sqrt{2\pi n_e}$, and $d_w^{-1} = a n_e$~\cite{wdw}.
Equation~\eqref{eq:mur} gives $\mu_{\rm RI} \sim 10^4$ at $n_e = 3$, which is more than 300 times larger than  experimental $\mu$, so the RI scattering is still irrelevant in this regime.
We note, however, that additional mechanisms of disorder in the doping layers may lead to an increase of RI scattering, as discussed in Sec.~V of  Ref.~\onlinecite{sammon:2018}, although quantitative understanding of these mechanisms is still lacking.

\emph{Interface roughness.}
\label{sec:surface_roughness}
Scattering on interface roughness originates from fluctuations of the ground state energy due to $1-2$ monolayer local variations of the quantum well width $w$.
For the infinite barrier height ($V \rightarrow \infty$), the corresponding transport scattering rate $\tau^{-1}_{\rm IR} \propto w^{-6}$ \cite{gold:1986,gold:1987} and such dependence was confirmed in narrow $w<10$ nm GaAs quantum wells with AlAs barriers \cite{sakaki:1987,kamburov:2016}. 
However, in Al$_{x}$Ga$_{1-x}$As/GaAs quantum wells the barrier height is significantly reduced ($V \approx 0.75 x$ eV for $x<0.45$), and fluctuations of the ground state energy are diminished due to finite penetration of the electron wave function into the barrier~\cite{gold:1989b,li:2005}.
As a result, $\tau^{-1}_{\rm IR}$ is considerably suppressed compared to the case $V \rightarrow \infty$, and its dependence on $w$ weakens \cite{gottinger:1988,li:2005,luhman:2007,kamburov:2016}.

We calculate the IR-limited scattering rates following the approach of Refs.~\onlinecite{gold:1989b,li:2005}. 
We use the barrier height for $x = 0.12$ and take into account the difference in the electron effective mass in the GaAs well ($m^{\star} = 0.067\,m_0$) and in the Al$_{0.12}$Ga$_{0.88}$As barriers ($m_B  = 0.067 + 0.083x = 0.077\,m_0$).
At the end we also impose a constraint $k_Fw = 3.9$ \citep{wdw}.

Using the correlator of local well width variations $\ev{\Delta(\vb{r}) \Delta(\vb{r}')} = \Delta^2 \exp(-\abs{\vb{r} - \vb{r}'}^2/\Lambda^2)$, where $\Delta$ is the roughness height and $\Lambda$ is the roughness lateral size, the scattering potential due to interface roughness is given by 
\begin{align}\label{eq:usq2}
    \ev{\abs{U_{\rm IR}(q)}^2} = \frac{\pi}{\epsilon^2(q)} \Delta^2 \Lambda^2 \qty(\pdv{E}{w})^2 e^{-q^2\Lambda^2/4}\,,
\end{align}
where $E$ is the ground state energy for the finite potential well.
Here, the dielectric function $\epsilon(q)$, see Eq.~\eqref{eq:df}, is computed with the form factor $F_c(qw)$, see Eq.~\eqref{eq:fcq}, using finite-potential-well wave function. (See Supplemental Material for details \citep{note:sm}).

\begin{figure}[t]
    \centering
    \includegraphics[width = \linewidth]{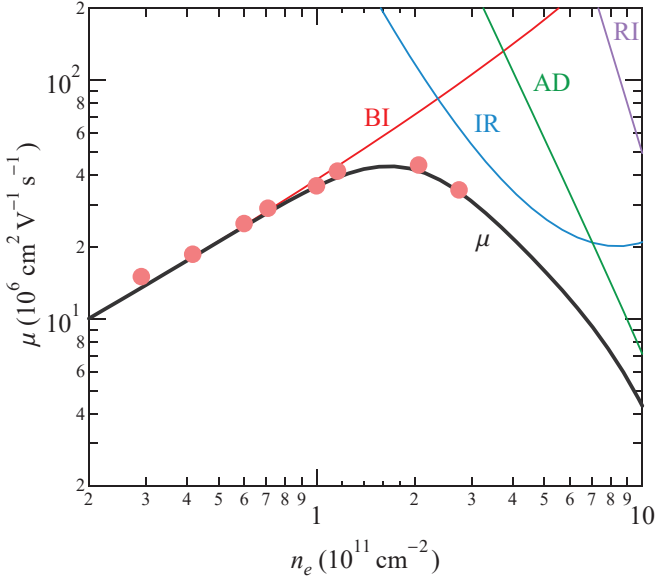}
    \caption{
    Mobility $\mu$ versus electron density $n_e$. 
   Circles are experimental data from Ref.~\onlinecite{chung:2021}. 
   Thin lines marked by BI, IR, AD, and RI represent $\mu_{\rm BI}$ ($N_1 = 1.4\times 10^{14}$ cm$^{-3}$, $N_2 = 0$), $\mu_{\rm IR}$ ($\Delta = 0.283$ nm, $\Lambda = 8$ nm, $x = 0.12$), $\mu_{\rm AD}$ ($x = 0.12)$, and $\mu_{\rm RI}$ [Eq.~\eqref{eq:mur}], respectively.
   Thick line represents $\mu$ limited by all contributions.
    }
    \label{fig:mobility_fit}
\end{figure}

By substituting Eq.~\eqref{eq:usq2} into Eq.~\eqref{eq:1_tau} we obtain the mobility due to interface roughness $\mu_{\rm IR}$ as shown in Fig.~\ref{fig:mobility_fit}.
In order to reproduce the experimental data at $n_e>2$, we arrived at $\Delta = 2.83$ {\AA} and $\Lambda = 80$ {\AA}. 
With these roughness parameters, $\mu_{\rm IR}$ becomes equal to $\mu_{\rm BI} (N_2=0)$ at $n_e \approx 2.4$ and, in the vicinity of this density, can be described by  $\mu_{\rm IR} \simeq 4.7 \times 10^2\,n_e^{-2}$.
We have also examined the effect of $x$ on the mobility limited by IR. 
By raising $x$ from 0.12 to 0.24, a value which is common for previous generation of samples, $\mu_{\rm IR}$ at $n_e = 3$ becomes smaller by 24\,\%, although this effect weakens at lower $n_e$.

\begin{figure}[t]
    \centering
    \includegraphics[width = \linewidth]{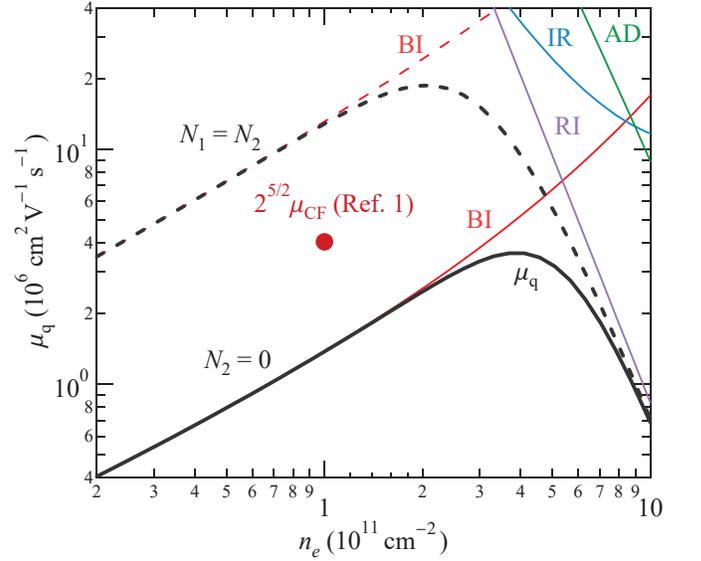}
    \caption{Quantum mobility $\mu_{\rm q}$ versus electron density $n_e$. 
   Thin solid lines marked by BI, IR, AD, and RI represent $\mu_{\rm q,BI}$ ($N_1 = 1.4\times 10^{14}$ cm$^{-3}$, $N_2 = 0$), $\mu_{\rm q,IR}$ ($\Delta = 0.283$ nm, $\Lambda = 8$ nm, $x = 0.12$), $\mu_{\rm q,AD}$ ($x = 0.12)$, and $\mu_{\rm q,RI}$ (Eq.~\eqref{eq:murq}), respectively.
   Thick solid line represents $\mu_{\rm q}$ limited by all contributions.
   Thick dashed line is $\mu_{\rm q}$ limited by all contributions but with $\mu_{\rm q,BI}$ computed for $N_1 = N_2 = 5 \times 10^{12}$ cm$^{-3}$.
   Solid circle shows quantum mobility obtained from the resistivity at filling factor $\nu=1/2$ measured in Ref.~\onlinecite{chung:2021}. 
   }
    \label{fig:mobility_q}
\end{figure}

\emph{Alloy disorder.}
Alloy disorder (AD) scatters electrons due to the wave function tails extending into the Al$_{x}$Ga$_{1-x}$As barriers. 
Usually, in typical high mobility samples, this scattering mechanism is deemed irrelevant, see e.g., Ref.~\onlinecite{li:2005,umansky:2013}.
However, in light of lower $x$ used in the new generation of samples \citep{chung:2021}, it is important to revisit this scattering source.
Following Refs.~\onlinecite{ando:1982,li:2005,gold:2013}, we write the scattering potential as 
\begin{equation}\label{eq:urm2}
    \ev{\abs{U_{\rm AD}(q)}^2} = \frac{(\Delta E_c)^2}{\epsilon^2(q)}\Omega x (1-x)  \int \limits_{\abs{z}>w/2} \abs{\psi(z)}^4 dz ,
\end{equation}
where $\Omega = a^3/4$, $a = 5.67$ $\si{\angstrom}$ is the lattice constant, and $\Delta E_c \simeq 1$ eV is the conduction band discontinuity at the $\Gamma$-point of GaAs/AlAs interface.
As for the case of IR, we use finite-potential-well wave function to calculate the form factor in $\epsilon(q)$ (see Supplementary Material \citep{note:sm}).
Substituting Eq.~\eqref{eq:urm2} into Eq.~\eqref{eq:1_tau} and applying the constraint $k_Fw = 3.9$ \citep{wdw}, we obtain the mobility limited by AD scattering $\mu_{\rm AD}$ as shown in Fig.~\ref{fig:mobility_fit}. 
We note that $\mu_{\rm AD}$ can be well described by $\mu_{\rm AD} \simeq 7.1 \times 10^3\, n_e^{-3}$ and that it becomes equal to $\mu_{\rm BI}$ at $n_e \approx 3.7$ and to $\mu_{\rm IR}$ at $n_e \approx 7$.
We thus conclude that alloy disorder scattering cannot be ignored at higher densities.
We have also examined the effect of $x$ on mobility limited by alloy disorder. 
By increasing $x$ from 0.12 to 0.24, $\mu_{\rm AD}$ increases by a factor of $2.4$ for $n_e < 10$.

We now turn to the effects of different scattering mechanisms on quantum mobility $\mu_{\rm q} = e \tau_{\rm q}/m^{\star}$, where the quantum scattering rate is given by
\begin{align}
	\frac{1}{\tau_{\rm q}} =\frac{2m^{\star}}{\pi \hbar^3}\int \limits_0^{2k_F}\frac{dq}{\sqrt{4k_F^2 - q^2}}\, \ev{\abs{U(q)}^2}. \label{eq:1_tauq}
\end{align}

\emph{Background impurities.}
Since background impurities far away from the quantum well contributes to $\mu_{\rm q}$ significantly, excess electron screening cannot be ignored and the scattering potential is no longer given by Eq.~\eqref{eq:u_i1_mu}.
As a very good approximation we can think of a perfect screening, such that the excess electron screening length is zero. 
However, the expression of scattering potential $U(q)$ in this approximation is still cumbersome so we leave it to Supplemental Material \citep{note:sm} as a reference for an interested reader.
As discussed above, two different impurity distributions, $N_2=0$ and $N_1=N_2$, can describe the experimental $\mu$ reasonably well at $n_e < 1$.
However, these distributions give very different values for $\mu_{\rm q}$, as shown in Fig.~\ref{fig:mobility_q}.
Indeed, we find that $\mu_{\rm q, BI}(N_2=0)$ (solid curve marked ``BI'') is an order of magnitude smaller than $\mu_{\rm q, BI}(N_1 = N_2)$ (dashed curve marked ``BI'').
More specifically, we find that $\mu_{\rm q, BI}(N_2=0) \simeq 1.4\,n_e^{0.85}$, whereas
$\mu_{\rm q, BI}(N_1=N_2) \simeq 13.3\,n_e^{0.87}$.
As a result, future experiments measuring quantum mobility should be able to distinguish between these two distributions.

\emph{Remote donors.}
For quantum mobility limited by RI scattering, we use Eq.~(23) from \rref{sammon:2018} valid at $1 - f = n_e/2n < 0.5$ for $n_e < 10$~\cite{remote}: 
\begin{align}
    \mu_{\rm q, RI} &= 6.5\,\frac{e}{\hbar} \frac{n k_F d_w^3}{n_e} = 2.6\times 10^3\, n_e^{-7/2}\, , \label{eq:murq}
\end{align}
where in the final expression we used $n=15$, $k_F = \sqrt{2\pi n_e}$, and $d_w^{-1} = a n_e$~\cite{wdw}.
Even though Eq.~\eqref{eq:murq} yields very large $\mu_{\rm q, RI} \simeq 56$ at $n_e = 3$, RI scattering is still expected to limit $\mu_{\rm q}$ at higher $n_e$, see Fig.~\ref{fig:mobility_q}.

\emph{Interface roughness and alloy disorder.}
Contributions from IR and AD can be calculated using Eqs.~\eqref{eq:usq2}, \eqref{eq:urm2}, and \eqref{eq:1_tauq}.
In particular we find that $\mu_{\rm q, IR} \simeq 1.2 \mu_{\rm IR}$ and $\mu_{\rm q, AD} \simeq 1.3 \mu_{\rm AD}$.
As shown in Fig.~\ref{fig:mobility_q}, these contributions are considerably smaller than the RI contribution at all relevant $n_e$.
As a result, one expects a crossover from BI-limited to RI-limited quantum mobility for either model of BI distribution.
The value and the position of the maximum at the quantum mobility crossover should yield information on the BI distribution.

Next, we comment on the relation between the quantum mobility and the mobility of composite fermions $\mu_{\rm CF}$ at filling factor $\nu = 1/2$.
By comparing the expression of the longitudinal resistivity at $\nu = 1/2$, Eq.\,(5.11) of Ref.~\onlinecite{halperin:1993},
\begin{align}\label{eq:rho1/2}
    \rho_{1/2} = \frac{n_i}{n_e} \frac{1}{k_F z} \frac{2\sqrt{2}\pi \hbar}{e^2},
\end{align}
and the expression for quantum mobility, Eq.~(6) in Ref.~\onlinecite{sammon:2018},
\begin{align}
    \mu_{\rm q} = \frac{2e}{\pi \hbar} \frac{k_F z}{n_i},
\end{align}
one can conclude that
\begin{align}\label{eq:muq_rho1/2}
    \mu_{\rm q} = \frac{4\sqrt{2}}{e n_e \rho_{1/2}},
\end{align}
which implies that $\mu_{\rm q}  = 4\sqrt{2} \mu_{\rm CF}$.
Here, $n_i$ is the 2D concentration of random impurities in a thin layer at a distance $z$ away from the center of the quantum well.
In order to obtain $\rho_{1/2}$ and $\mu_{\rm q}^{-1}$ for impurities with 3D concentration $N(z)$, one should replace $n_i$ by $N(z) dz$ and integrate over $z$~\footnote{To avoid the divergence at $z=0$, one can average $z$ in the denominator using $\ev{z} = \int dz' \abs{\psi(z')}^2 \abs{z-z'}$, where $\psi(z)$ is the wave function along $z$ direction.}. 
This integration does not change the relation Eq.~\eqref{eq:muq_rho1/2} between $\mu_q$ and $\rho_{1/2}$, so that it holds for both BI and RI scattering~\footnote{On the other hand, AD and IR do not change the 2D electron concentration inside the quantum well, so they do not contribute to the gauge field fluctuations and $\rho_{1/2}$.}. 

From the experimental data in Ref.~\onlinecite{chung:2021}, the longitudinal resistance is $R_{1/2} = 35$ $\si{\ohm}$ at $\nu = 1/2$ and $n_e = 1$. 
Using the geometry of experiments in Ref.~\onlinecite{chung:2021}, we estimate $\rho_{1/2} \approx 2.5 R_{1/2}$.
As a result, the quantum mobility at $n_e = 1$ is estimated as $\mu_{\rm q} = 4.0$.
This data point, filled circle in Fig.~\ref{fig:mobility_q}, falls in between our predictions for $N_2 = 0$ and $N_1=N_2$.
In order to fit this point, while keeping $\mu$ the same as shown in Figs.~\ref{fig:mobility_low_ne} and \ref{fig:mobility_fit}, we need $N_1 \approx 4 \times 10^{13}$ cm$^{-3}$ and $N_2 \approx 4 \times 10^{12}$ cm$^{-3}$.

Finally, we mention that hydrodynamics \citep{alekseev:2016} or scattering on oval defects \citep{bockhorn:2014} might affect the zero-field resistivity and, as a result, the inferred mobility.
Both mechanisms manifest as negative magnetoresistance  in weak magnetic field \citep{dai:2010,hatke:2011b,bockhorn:2011,hatke:2012a,shi:2014a,shi:2014c,bockhorn:2014} which can also be seen in Fig.\,3(b) of \rref{chung:2021}.
However, since no experimental studies of this magnetoresistance are yet available, we cannot comment on its origin.

In summary, we have examined roles of different scattering sources on the transport and quantum mobilities in the new generation of ultrahigh mobility GaAs/Al$_{0.12}$As$_{0.88}$ quantum wells \citep{chung:2021}.
While at lower electron densities both mobilities are limited by background impurity scattering, interface roughness (remote impurity) scattering is the likely source limiting transport (quantum) mobility at large electron densities.
Our predictions for quantum mobility are in agreement with the value estimated from the mobility of the composite fermions at filling factor $\nu= 1/2$ in a sample with $n_e = 1 \times 10^{11}$ cm$^{-2}$ \cite{chung:2021}. 
Future measurements of quantum mobility should provide insight on the distribution of background impurities in the GaAs quantum well and in the AlGaAs barriers.

\begin{acknowledgments}
We thank L. N. Pfeiffer and M. Shayegan for discussions and clarifying the experimental details.
Y.H. was partially supported by the William I. Fine Theoretical Physics Institute.
M.A.Z. acknowledges support by the U.S. Department of Energy, Office of Science, Basic Energy Sciences, under Award No. ER 46640-SC0002567. 
\end{acknowledgments}

\medskip
%

\setcounter{equation}{0}
\setcounter{figure}{0}
\setcounter{table}{0}
\setcounter{page}{1}
\makeatletter
\renewcommand{\theequation}{S\arabic{equation}}
\renewcommand{\thefigure}{S\arabic{figure}}
\section{Supplementary materials}
\subsection{Well width $w$ and setback $d_w$ versus electron density}
\begin{figure}[b]
\includegraphics{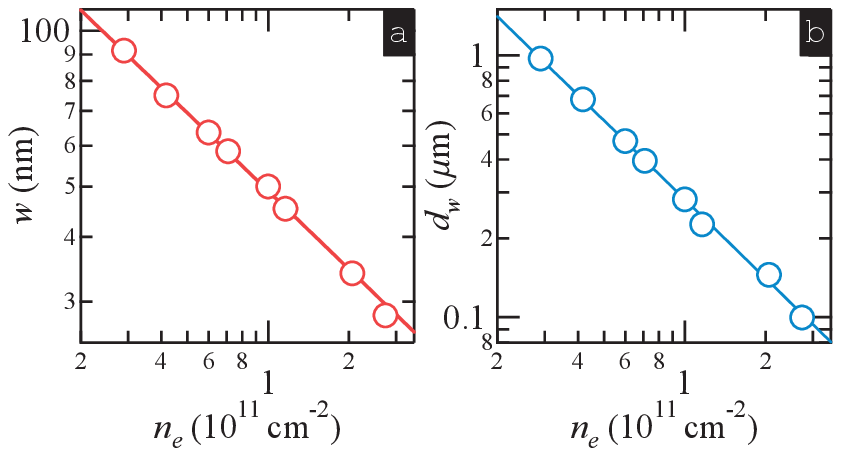}
\vspace{-0.05 in}
\caption{
(a) Quantum well width $w$, and (b) effective setback distance $d_w =  d + w/2$ as a function of the carrier density $n_e$ in a new generation, doping-well samples from \rref{chung:2021}.
Solid lines are power-law fits yielding $w = 49\,n_e^{-1/2}$ and $d_w = 0.28\, n_e^{-1}$, in units as marked.
These fits correspond to constraints $k_F w = 3.9$
 and 
$d_w^{-1} =  a n_e$, where $a = 3.55$ nm.
}
\vspace{-0.15 in}
\label{fig0}
\end{figure}

To obtain the constraints on the quantum well width $w$ and on the effective spacer width $d_w = d + w/2$ we have fitted the experimental data of \rref{chung:2021} as illustrated in Fig.~\ref{fig0}.
Open circles in panel (a) and (b) of Fig.~\ref{fig0} represent $w$ and $d_w$, respectively, as a function of electron density $n_e$.
Solid lines are power law fits to the data, yielding $w = 49\,n_e^{-1/2}$ and $d_w = 0.28\,n_e^{-1}$, in units as marked.
These relationships are equivalent to $k_F w = 3.9$, where  $k_F = \sqrt{2\pi n_e}$ is the Fermi wave number, and $d_w^{-1} =  a n_e$, where $a = 3.55$ nm, used in the main text.
Here, the first condition corresponds to a maximum electron density at which the second subband of the GaAs quantum well still remains unoccupied, whereas the second condition is governed by electrostatics of the device.

\label{sec:wdw}

\subsection{Scattering potential of a single background impurity}
\label{sec:potential}
We derive the screened potential of a single charged impurity at a distance $z$ away from the center of a quantum well of thickness $w$ using random phase approximation (RPA) as in Ref.~\onlinecite{sammon:2018}.
There are two sources of screening for background impurities: they can be screened by 2DEG in the quantum well or by excess electrons (EEs) remaining in the remote doping wells separated from the center of the 2DEG by distance $d_w =d +w/2$.
We introduce the notations $U_{{\rm i}, k}$ for the screened impurity interaction and ${\tilde U}_{{\rm i},k}$ for the unscreened impurity interaction, where the subscript i denotes the impurity and subscript $k$ can be equal to 1 (2DEG) or 2 (EEs). [Used below subscript $l$ can be equal to 1 (2DEG) or 2 (EEs) as well].
Following 
Ref.~\onlinecite{sammon:2018}, we obtain the screened Coulomb interaction between an impurity and an electron in the 2DEG,
\begin{align}\label{eq:u_i1}
    U_{{\rm i},1} =
    \frac{{\tilde U}_{{\rm i},1} (1 - \Pi_2 {\tilde U}_{2,2}) + {\tilde U}_{{\rm i},2} \Pi_2 {\tilde U}_{2,1}}{(1 - \Pi_1 {\tilde U}_{1,1})(1 - \Pi_2 {\tilde U}_{2,2}) - \Pi_1 \Pi_2 {\tilde U}_{1,2}^2}\,,
\end{align}
where bare interactions ${\tilde U}_{{\rm i}, k}$ and ${\tilde U}_{k,l}$ are given by
\begin{gather}
    {\tilde U}_{{\rm i}, k}(q,z) = \frac{2\pi e^2}{\kappa q} \int \limits_{-\infty}^{+\infty} dz' \;\lambda_k(z') e^{-q\abs{z- z'}}\,, \\
    {\tilde U}_{k,l}(q) = \frac{2\pi e^2}{\kappa q} \iint \limits_{-\infty}^{+\infty} dz \;dz' \; \lambda_k(z) \lambda_l(z') e^{-q\abs{z-z'}}\,.
\end{gather}
Here, $\lambda_1(z) = \abs{\psi(z)}^2$ and $\lambda_2(z) = \delta(\abs{z}-d_w)$ are the linear densities of the 2DEG and of the EEs respectively, and the polarization functions $\Pi_1 = -(\kappa q_{\rm TF}/2\pi e^2) [1 - G(h)]$ and $\Pi_2 = - (\kappa r_s^{-1} / 2\pi e^2) [1 - G(h)]$ are calculated using Thomas-Fermi approximation with local field corrections.

Within infinite-potential-well approximation, $\lambda_1(z) = (2/w) \cos^2(\pi z/w) \Theta(w/2 - \abs{z})$, we can write down the analytical expressions for bare interactions.
For ${\tilde U}_{{\rm i}, 1}$, we find
\begin{align}
    {\tilde U}_{{\rm i}, 1}(q,z) = \frac{2\pi e^2}{\kappa q}
    \begin{cases}
        F_0(qw) e^{-q\abs{z}}, & \abs{z} > w/2\,, \\
        G_0(q,z), & \abs{z} < w/2\,,
    \end{cases}
\end{align}
where 
\begin{align}
    F_0(x) = \frac{8 \pi^2}{x(x^2 + 4\pi^2)} \sinh(x/2)\, ,
\end{align}
and
\begin{align}
    G_0(q,z) &= F_0(qw) {\rm csch}(qw/2) (1-e^{-qw/2}\cosh(qz)) \nonumber\\
    &+ \frac{4qw\cos^2(\pi z / w)}{4\pi^2 + q^2 w^2}\,.
\end{align}

For ${\tilde U}_{{\rm i}, 2}$, we have
\begin{align}
    {\tilde U}_{{\rm i}, 2} = \frac{2\pi e^2}{\kappa q} (e^{-q\abs{z-d_w}} + e^{-q\abs{z+d_w}})
\end{align}

The electron-electron interactions are given by
\begin{align}
    {\tilde U}_{1, 1} = \frac{2\pi e^2}{\kappa q} F_c(qw),
\end{align}
\begin{align}
    {\tilde U}_{1, 2} = {\tilde U}_{2, 1} = \frac{2\pi e^2}{\kappa q} 2e^{-qd_w}F_0(qw),
\end{align}
\begin{align}
    {\tilde U}_{2, 2} = \frac{2\pi e^2}{\kappa q} (2+2e^{-2qd_w}),
\end{align}
where
\begin{align}
    F_c(x) = \frac{20 \pi^2 x^3 + 3 x^5 - 32 \pi^4 (1-e^{-x} - x)}{x^2(4\pi^2 + x^2)^2}.
\end{align}

Screening by EEs affects only the potential of impurities located in the vicinity of the doping wells. 
These impurities practically do not contribute to momentum relaxation, so we can calculate the mobility assuming  $\Pi_2 = 0$. 
The corresponding $U_{i,1}$ in Eq.~\eqref{eq:u_i1} with $\Pi_2 = 0$ can be simplified as 
\begin{align}
    U_{i,1} = \frac{{\tilde U}_{{\rm i}, 1}}{1-\Pi_1 {\tilde U}_{1, 1}},
\end{align}
which is equal to $U_1$ defined in Eq.~(3) of the main text.
However, for quantum mobility EE screening plays an important role. 
The screening radius of EEs is given by $r_s\simeq 0.5 n^{-1/2}$~\cite{sammon:2018a}, valid for large filling fraction of remote donors $f = 1 - n_e/2n > 0.5$ ~\footnote{
Since background impurities scattering dominates at $n_e < 3 \ll n$, $1-f < 0.1 \ll 1$ and we can use $r_s\simeq 0.5 n^{-1/2}$ from Ref.~\onlinecite{sammon:2018a}. },
where $n_e$ is the 2DEG density and $n= 1.5 \times 10^{12}$ cm$^{-2}$ is the concentration of donors. %
We have confirmed that our results for $\mu_{\rm q}$ with $r_s\simeq 0.5 n^{-1/2} \approx 4$ nm are practically the same as for $r_s =0$ or $\Pi_2 = \infty$.

\subsection{Finite-potential-well solution}
\label{sec:finite}
The Schr{\"o}dinger equation is given by
\begin{equation}
    - \frac{\hbar^2}{2 m(z)} \dv[2]{\psi(z)}{z} + V(z) \psi(z) = E \psi(z)\,, 
\end{equation}
where $m(z) = m^\star = 0.067\,m_0$ at $\abs{z} < w/2$ and $m(z) = m_B = (0.067 + 0.083x)m_0 = 0.077\,m_0$ at $\abs{z} > w/2$, and the effective finite potential well is described by
\begin{equation}
    V(z) = 
    \begin{cases}
        0\qc \, & \abs{z} < w/2\,,\\
        V_0 - E_F\qc \, & \abs{z} > w/2\,,
    \end{cases}
\end{equation}
where $V_0 = 90$ meV ($x = 0.12$) and $E_F = \hbar^2 k_F^2/2m^{\star}$ is the Fermi energy.
Below we use $V$ to represent $V_0 - E_F$ for simplicity.
The bound state solutions ($E<V$) for the lowest subband are
\begin{equation}\label{eq:eigenfunction}
\psi(z) =
    \begin{cases}
        C e^{-\eta \abs{z}}, \, & \abs{z}>w/2\,, \\
        D \cos(kz), \, & \abs{z}<w/2\,, \\
    \end{cases}
\end{equation}
where $k = \sqrt{2m^{\star} E/\hbar^2}$, $\eta = \sqrt{2 m_B (V - E)/\hbar^2}$.
At $z=\pm w/2$, $\psi(z)$ is continuous while $\psi'(z)$ has a finite discontinuity because of the difference in effective masses: 
\begin{gather}
    C e^{-\eta w/2} = D \cos(kw/2)\,, \label{eq:help}\\
    \frac{C}{m_B} \eta e^{-\eta w/2} = \frac{D}{m^{\star}} k \sin(kw/2)\,.\label{eq:help2}
\end{gather}
The normalization of $\psi(z)$ gives
\begin{equation}
    C^2 \int_{w/2}^{\infty} e^{-2\eta z} dz + D^2 \int_0^{w/2} \cos^2(kz) dz = 1/2,
\end{equation}
Combining with Eq.~\eqref{eq:help}, one has
\begin{equation}
    C = \qty[\frac{1}{\eta} e^{-\eta w} + \frac{kw + \sin(kw)}{2k \cos^2(kw/2)} e^{-\eta w}]^{-1/2}. \label{eq:C}
\end{equation}

By dividing Eq.~\eqref{eq:help2} by Eq.~\eqref{eq:help}, one obtains
\begin{equation}\label{eq:bound_state}
    \tan(kw/2) = \frac{m^{\star}}{m_B} \frac{\eta}{k}\,.
\end{equation}
Introducing dimensionless quantities $\tilde E = E/E_0$, $\tilde V = V/E_0$, where $E_0 = \hbar^2 \pi^2/2 m^{\star} w^2$ is the ground state energy for infinite well, 
in Eq.~\eqref{eq:bound_state} we arrive at~\footnote{See for example, Eq.~(5c) in Ref.~\onlinecite{gold:1989b} or Section 4.9 in Ref.~\onlinecite{davies:1997}.}
\begin{equation}\label{eq:energy}
    \tilde E + \frac{m_B}{m^{\star}} \tilde E \tan^2{\qty(\sqrt{\tilde E} \pi /2)} = \tilde V\,.
\end{equation}

If $\tilde V \to \infty$ and $m_B = m^{\star}$, then $\tilde E = 1$. For any finite $\tilde V$, we have $0< \tilde E < 1$.
For example, at small electron concentration $n_e < 1 \times 10^{11}$ cm$^{-2}$ with $k_F w = 3.9$, we have $\tilde V > 37 \gg 1$, which justifies our infinite-potential-well approximation at lower $n_e$.
On the other hand, at large electron concentration $n_e \in (5,15) \times 10^{11}$ cm$^{-2}$, $\tilde V$ decreased significantly from 6.2 to 1.1, and therefore finite-potential-well effect such that $0< \tilde E < 1$ has to be taken into account.

The local energy fluctuation due to interface roughness is given by
\begin{equation}
    \delta E (\vb{r}) = \pdv{E}{w} \Delta(\vb{r})\,,
\end{equation}
where $\Delta(\vb{r})$ is the local variation of the well width at a position $\vb{r}$.
Taking derivative with respect to $w$ for both sides in Eq.~\eqref{eq:energy}, we obtain the expression for $\pdv*{E}{w}$
\begin{equation}\label{eq:dEdw}
    \pdv{E}{w} = - \frac{2E}{w} \qty{1 + g\qty(\frac{m_B}{m^{\star}}, \frac{E}{V}) {\tilde V}^{-1/2} }^{-1}\,,
\end{equation}
and 
\begin{equation}
    g(x,y) =\frac{2}{\pi}\qty[x^{-1/2} + y(x^{1/2} - x^{-1/2})]^{-1}(1 - y)^{-1/2}\,.
\end{equation}
Substituting Eq.~\eqref{eq:dEdw} in Eq.~(8) in the main text, we obtain the scattering potential for the interface roughness.

\end{document}